\documentclass[]{article}
\usepackage{graphicx,enumerate,verbatim,bbold}
\usepackage{amsmath,amssymb,amsthm,float,mathrsfs}
\usepackage[affil-it]{authblk}
\usepackage[colorlinks]{hyperref}
\usepackage[T1]{fontenc}
\usepackage[utf8]{inputenc}
\usepackage{lmodern}
\usepackage{xspace}
\usepackage{siunitx}
\usepackage{caption}
\usepackage{pgfplots,tikz,subcaption}

\usepackage{pdfcomment}

\newcommand{\braket}[2]{\mbox{$ \langle #1 | #2 \rangle $}}

\newcommand{\ket}[1]{\mbox{$ | #1 \rangle $}}

\newcommand{\com}[2]{\left[ #1\,,\,#2 \right]}

\newcommand{\w}{\omega}
\newcommand{\sig}{\sigma}

\newcommand{\one}{\leavevmode\hbox{\small1\normalsize\kern-.33em1}}%

\newcommand{\Tr}{\mbox{Tr}}

\newcommand{\dagg}{^{\dag}}
\newcommand{\inv}[1]{\frac{1}{#1}}
\newcommand{\nn}{\nonumber}
\newcommand{\proj}[2]{\mbox{$ |#1 \rangle \langle #2 | $}}
\newcommand{\eg}{\textit{e.g.}\@\xspace}
\newcommand{\ie}{\textit{i.e.}\@\xspace}

\newcommand{\half}[1]{\frac{#1}{2}}
\newcommand{\pare}[1]{\left( #1 \right)}
\newcommand{\abs}[1]{\left|#1\right|}
\newcommand{\mean}[1]{\left\langle#1\right\rangle}
\newcommand{\chsh}{\mathcal{B}}
\newcommand{\ketbra}[2]{\left|#1\middle\rangle\middle\langle#2\right|}
\DeclareMathOperator{\dint}{d\!}

\def\beq{\begin{equation}}
\def\eeq{\end{equation}}
\def\beqa{\begin{eqnarray}}
\def\eeqa{\end{eqnarray}}
\newcommand{\eq}[1]{(#1)}
\newcommand{\fig}[1]{Fig.~#1}
\newcommand{\dani}[1]{{\color{black} #1}}

\newcommand{\colin}[1]{{\color{black}#1}}
\newcommand{\comm}[1]{}
\newcommand{\subfigref}[2][{}]{\hyperref[#2]{\ref{#2}#1}}
\DeclareCaptionLabelFormat{andtable}{#1~#2  \&  \tablename~\thetable}

\newcommand{\col}[1]{\color{black} #1}

\title{Realistic loophole-free Bell test with atom-photon entanglement}
\author[1]{C. Teo}
\author[2,5]{M. Araújo}
\author[2,6]{M. T. Quintino}
\author[1]{J. Min\'{a}\v{r}}
\author[1,7]{D. Cavalcanti}
\author[1,3]{V. Scarani\footnote{Electronic address: physv@nus.edu.sg}}
\author[4]{M. Terra Cunha}
\author[2]{M. Fran\c{c}a Santos\footnote{Electronic address: msantos@fisica.ufmg.br}}
\affil[1]{Centre for Quantum Technologies, National University of Singapore, Singapore}
\affil[2]{Departamento de F\'{i}sica, Universidade Federal de Minas Gerais, Caixa Postal 702, 30123-970 Belo Horizonte, MG, Brazil}
\affil[3]{Department of Physics, National University of Singapore, Singapore}
\affil[4]{Departamento de Matem\'{a}tica, Universidade Federal de Minas Gerais, Caixa Postal 702, 30123-970 Belo Horizonte, MG, Brazil}
\affil[5]{Faculty of
Physics, University of Vienna, Boltzmanngasse 5, 1090 Vienna, Austria}
\affil[6]{Département de Physique Théorique, Université de Genève, 1211 Genève, Switzerland}
\affil[7]{ICFO-Institut de Ciencies Fotoniques, 08860 Castelldefels (Barcelona), Spain}

\begin{document}
\maketitle

\begin{abstract}
 The establishment of nonlocal correlations, guaranteed through the violation of a Bell inequality, is not only important from a fundamental point of view, but constitutes the basis for device-independent quantum information technologies. Although several nonlocality tests have been performed so far, all of them suffered from either the locality or the detection loopholes. \dani{Among the proposals to overcome these problems are the use of atom-photon entanglement and hybrid photonic measurements (\textit{eg.} photo-detection and homodyning).
 Recent studies have suggested that the use of atom-photon entanglement can lead to Bell inequality violations with moderate transmission and detection efficiencies.
 Here
 we  combine these ideas and  propose an experimental setup realizing a simple atom-photon entangled state that
  can be used to obtain nonlocality when considering realistic experimental parameters including detection efficiencies and losses due to required propagation distances.}
\end{abstract}

\section*{Introduction}

In recent years several applications of quantum nonlocality have been proposed \cite{diqkd,dirandom,distateest,rabelo2011device} (see a recent review on Bell nonlocality \cite{review}). These proposals are based on the fact that nonlocal correlations can be certified without any assumption on the internal mechanisms of the devices used in the experiment. Thus, once established, nonlocal correlations can be used in what is now referred as, {device-independent} protocols.

Nonlocal correlations can be obtained by measuring entangled quantum systems in appropriately chosen local observables. This is called a Bell test since the nonlocal nature of the measurement outcomes can be certified by the violation of certain constraints known as Bell inequalities \cite{review}. Many Bell tests have been performed in the last few decades, but no nonlocal correlations have been strictly established so far. This is because all of the performed experiments suffered either from the detection loophole or the locality loophole \cite{review}.

\dani{Experiments using entangled photons have reported Bell inequality violations closing separetly the locality \cite{aspect_experimental_1982,weihs_violation_1998,scheidl_violation_2010} and the detection \cite{zeilinger12} loopholes. On the other hand, the detection loophole has been closed with stationary systems like atoms, ions and circuits \cite{matsukevich_bell_2008, rowe_experimental_2001, ansmann_violation_2009}. The main technological challenge to close both loopholes simultaneously is to have both efficient detection long-distance entanglement.}


\dani{In this work, we propose an experimental setup involving available technology to implement a loophole-free Bell test.
It uses a single atom coupled to the field of a cavity in order to produce a specific entangled state between the atom and the light emitted by the cavity
and combines efficient detection schemes on the atomic side and the coherent nature of the light field to perform hybrid  detection on the photonic side \cite{cavalcanti10,quintino12}. Our scheme considers experimental effects neglected in previous proposals \cite{araujo11,sangouard11} and thus puts current atom-photon systems as good candidates to demonstrate loophole-free nonlocal correlations}


The paper is structured as follows: first, we introduce the target state and the measurement settings used in the Bell test. We then give an explanation of how this state can be produced by means of cavity quantum electrodynamics (QED) techniques and discuss the relevant and feasible parameter regime which yields the desired state. We also discuss an implementation of our scheme in optical cavities, and optimize the \colin{Clauser-Horne-Shimony-Holt (CHSH) inequality \cite{chsh69}} violation for currently available experimental parameters. Finally, we discuss possible circuit QED implementations in the final section and compare our proposal to previous ones.
\section*{Results}
\subsection*{The ideal case}
\label{sec:ideal}

Our proposal takes advantage of two facts first noticed in \cite{brunner07} and \cite{cabello07}. First, highly efficient detections are typically available for the electronic levels of single atoms \cite{henkel2010highly}. At the same time, photons propagating either in free space or in low loss optical fibers are excellent candidates for carriers of information over long distances. \dani{Furthermore, the combination of high efficient homodyne detection with photodetection 
greatly reduces the} \colin{required minimum} \dani{photodetection efficiency \cite{cavalcanti10,quintino12, araujo11,sangouard11,Brask12}.}

Our starting point is to consider a state of the form
\begin{equation}
		\ket{\psi_\alpha} = \cos\nu\ket{s,0}+\sin\nu\ket{g,\alpha}, \label{eqn:target_state}
\end{equation}
where $\ket{g}$, $\ket{s}$ are two atomic states and $\ket{0}$ and $\ket{\alpha}$ denote states of the electromagnetic field (vacuum and a coherent state respectively) with
 \begin{equation}
   \ket{\alpha} := e^{-\frac{\abs{\alpha}^2}{2}}\sum_{n=0}^\infty \frac{\alpha^n}{\sqrt{n!}}\ket{n},
 \end{equation}
 where $\ket{n}$ are the energy eigenstates of the field. As we are going to see, this state violates a Bell inequality even for low efficiency photodetection and can be well approximated with current technology.
 Although we motivated this work by considering a real atom, the state \eq{\ref{eqn:target_state}} can be in principle devised using different platforms, \eg, superconducting qubits, quantum dots, nitrogen-vacancy center and equivalent systems. 
 We consider the \colin{CHSH} Bell inequality \cite{chsh69} 
 which states that the correlations obtained are nonlocal if the following inequality is violated:
 \begin{equation}
\mean{\chsh}= \Tr(\rho\chsh)\leq2,
 \end{equation}
	where
	$\rho$ is the quantum state under scrutiny and $\chsh$ denotes the Bell operator
\begin{equation}
  \chsh=A_0\otimes B_0 + A_0\otimes B_1 + A_1\otimes B_0 -A_1\otimes B_1,
\end{equation}
	with $A_i$ and $B_j$ being observables with outcomes $\pm1$.
	Here, we set the atomic observables as
\begin{equation}
A_0 = \cos\gamma \sigma_z + \sin\gamma \sigma_x , \quad A_1= \cos\gamma \sigma_z - \sin\gamma \sigma_x,
\end{equation}
where $\sigma_i$ are the usual Pauli matrices.
	In the photonic part we consider dichotomized photodetection and $X$ quadrature operators \dani{(homodyne detection)}\cite{cavalcanti10,quintino12}:
	\begin{equation}
	 B_0=2\ketbra{0}{0} - \one,  \quad B_1= 2\int_{-b}^b \dint x\,\ketbra{x}{x} - \one,
	\end{equation}
	where $\ket{x}$ is the eigenstate of the quadrature operator $X = \frac{a + a\dagg}{\sqrt{2}}$, and $a$ is the annihilation operator of the field.
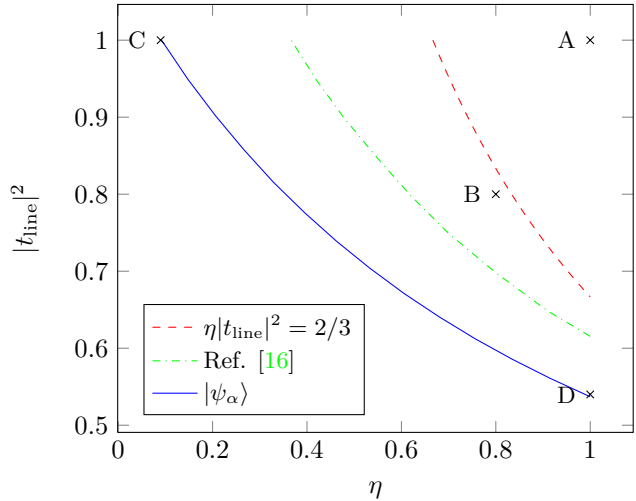
\begin{figure}[ht]
	\centering
	\begin{tikzpicture}
	\begin{axis}[
	xlabel=$\eta$,
	ylabel=$|t_{\rm line}|^2$,
	legend style={anchor=north west,cells={anchor=west},at={(0.05,.3)}, font = \small}
	]

	\addplot[color=red,domain=0.6666:1,dashed]{2/(3*x)};
	\addlegendentry{$\eta |t_{\rm line}|^2 = 2/3$}

	\addplot[color=green,mark=none,dashdotted] coordinates
	{(1, 0.615395) (0.91, 0.64827) (0.8, 0.698019) (0.7, 0.749074) (0.62, 0.797871) (0.55, 0.846303) (0.477214, 0.9) (0.42, 0.948192) (0.366631, 1)}; \addlegendentry{Ref. \cite{sangouard11}}

	\addplot[color=blue,mark=none] coordinates
{(1,0.53701)(0.99631,0.538)(0.91466,0.561)(0.834,0.58592)(0.756,0.61234)(0.67932,0.641)(0.605,0.67133)(0.53292,0.704)
(0.46235,0.739)(0.395,0.776)(0.329,0.81544)(0.266,0.858)(0.206,0.90201)(0.148,0.94901)(0.093,0.999)(0.092001,1)}; \addlegendentry{$\ket{\psi_\alpha}$}

	\addplot[only marks,color=black,mark=x] coordinates
	{(0.09, 1) (1, 0.54) (.8, .8) (1, 1)};
\node[color=black,font=\small] at (axis cs:0.95, 1) {A};
\node[color=black,font=\small] at (axis cs:0.75, 0.8) {B};
\node[color=black,font=\small] at (axis cs:0.04, 1) {C};
\node[color=black,font=\small] at (axis cs:0.95, 0.54) {D};
	\end{axis}
	\end{tikzpicture}
	\caption{
\textbf{Critical line above which \dani{nonlocal correlations can be obtained.}} The blue line corresponds to the ideal state of our proposal. The parameters $\alpha, \gamma, \nu, b$ were optimized for each point. For comparison, we include the curve $\eta |t_{\rm line}|^2 = 2/3$ (red) that results from the Eberhard bound \cite{eberhard_93,larsson01}, and the curve from the best experimental proposal to date in Ref \cite{sangouard11} (green). For the sake of illustration, we give the specific numbers for the points represented by crosses \colin{A, B, C and D in Table \ref{tab:Ideal_pts}.}} \label{comparison}
\end{figure}
\begin{table}[h!]
\centering
\begin{tabular}[b]{*{5}{|c}|}
\hline
 & A & B &C & D\\
 \hline
  $|t_{\rm line}|^2$ & 1 & 0.8 & 1 & 0.55 \\
 \hline
  $\eta$ &1 & 0.8 & 0.15 & 1 \\
  \hline
 $|\alpha|$ & 2.1 & 2.33 & 3.35 & 2.38\\
 \hline
 $\gamma$ & 0.55 & 0.34 & 0.14 & 0.03 \\
 \hline
 $\nu$ & 0.77 & 0.66 & 0.16 & 0.33\\
 \hline
 $b$ & 0.53 & 0.53 & 0.34 & 0.44 \\
 \hline
  $\mean{\chsh}$ & 2.32 & 2.07& $2^+$ & $2^+$\\
   \hline
\end{tabular}
\caption{{\bf Optimized parameters for 4 points on \fig{\ref{comparison}}.} The state and measurement parameters, $|\alpha|,\nu,\gamma$ and $b$, are each numerically optimized, given some detector efficiency $\eta$, and some transmission $|t_{\rm line}|^2$.} \label{tab:Ideal_pts}
\end{table}
	
\dani{Since quadrature measurements can be made very efficient (nearly perfect), there will be an asymmetry in the total efficiency of the photonic measurements: the total homodyning efficiency will be basically determined by transmission losses, while the total photodetection efficiency will be composed by transmission and typical photodetector efficiency. We thus model the transmission losses with a beam splitter with transmittance $t_{\rm line}$ (affecting both the photodetector and the homodyning apparatus) and the photodetector intrinsic efficiency as a beam splitter with transmittance $\sqrt{\eta}$ followed by a perfect detector.}


\fig{\ref{comparison}} summarizes the results. In ideal conditions (\ie point A in Table \ref{tab:Ideal_pts}), the CHSH value can reach $\mean{\chsh}=2.32$ for $|\alpha|=2.1$, which translates to an average of 4.41 photons in the coherent state. Moreover, a violation can be found even for $\eta=0.15$ with $|t_{\rm line}|^2=1$ (Point C in Table \ref{tab:Ideal_pts}), or transmittance $|t_{\rm line}|^2=0.55$ with $\eta = 1$ (Point D in Table \ref{tab:Ideal_pts}). Also, for a detection efficiency of $\eta =0.8$ and a transmission of $|t_{\rm line}|^2 = 0.8$ we find  a CHSH value of $2.07$ (Point B in Table \ref{tab:Ideal_pts}).
For comparison, we have also included the curve $\eta|t_{\rm line}|^2 = 2/3$ that results from the Eberhard bound \cite{eberhard_93,larsson01} , which is the required efficiency and transmission to perform a loophole-free experiment with photon polarization, and the curve from the best experimental proposal to date involving an atom and a photonic mode \cite{sangouard11}.

\subsection*{Realistic scenario} \label{sec:real_case}

We now show a scheme aiming to produce the state \eqref{eqn:target_state} in a cavity QED scenario.
The first part of the scheme \comm{realizing the state \eq{\ref{eqn:target_state}}}is depicted in \fig{\subfigref[a]{fig:cavity}}. An input field is incident on a cavity with a single three level atom prepared initially in a superposition of the states $g$ and $s$: $\ket{\psi}_{\rm atom} = \cos{\nu}\ket{s} + \sin{\nu}\ket{g}$. Fig.~\subfigref[b]{fig:cavity} shows the level structure of the atom. Transition $\ket{g}-\ket{e}$ is coupled dispersively to the cavity with detuning $\Delta = \omega_{\rm ge} - \omega_{\rm c}$, where $\omega_{\rm c}$ is the cavity frequency and $\omega_{\rm ge}$ is the frequency of the transition $\ket{g}-\ket{e}$. The cavity is asymmetric with decay rates of the mirrors $\kappa_{\rm b} \ll \kappa_{\rm c}$ and the cavity decay rate $\kappa = \kappa_{\rm b} + \kappa_{\rm c}$. This means that the field leaks out of the cavity essentially only through the right mirror (cf.\ Fig.~\subfigref[a]{fig:cavity}). The fact that one needs an asymmetric cavity is not \textit{a priori} obvious and will be explained below. We assume that the level $\ket{s}$ is detuned far enough such that it does not interact with the cavity field.

\begin{figure}[ht!]
\includegraphics[width = \columnwidth]{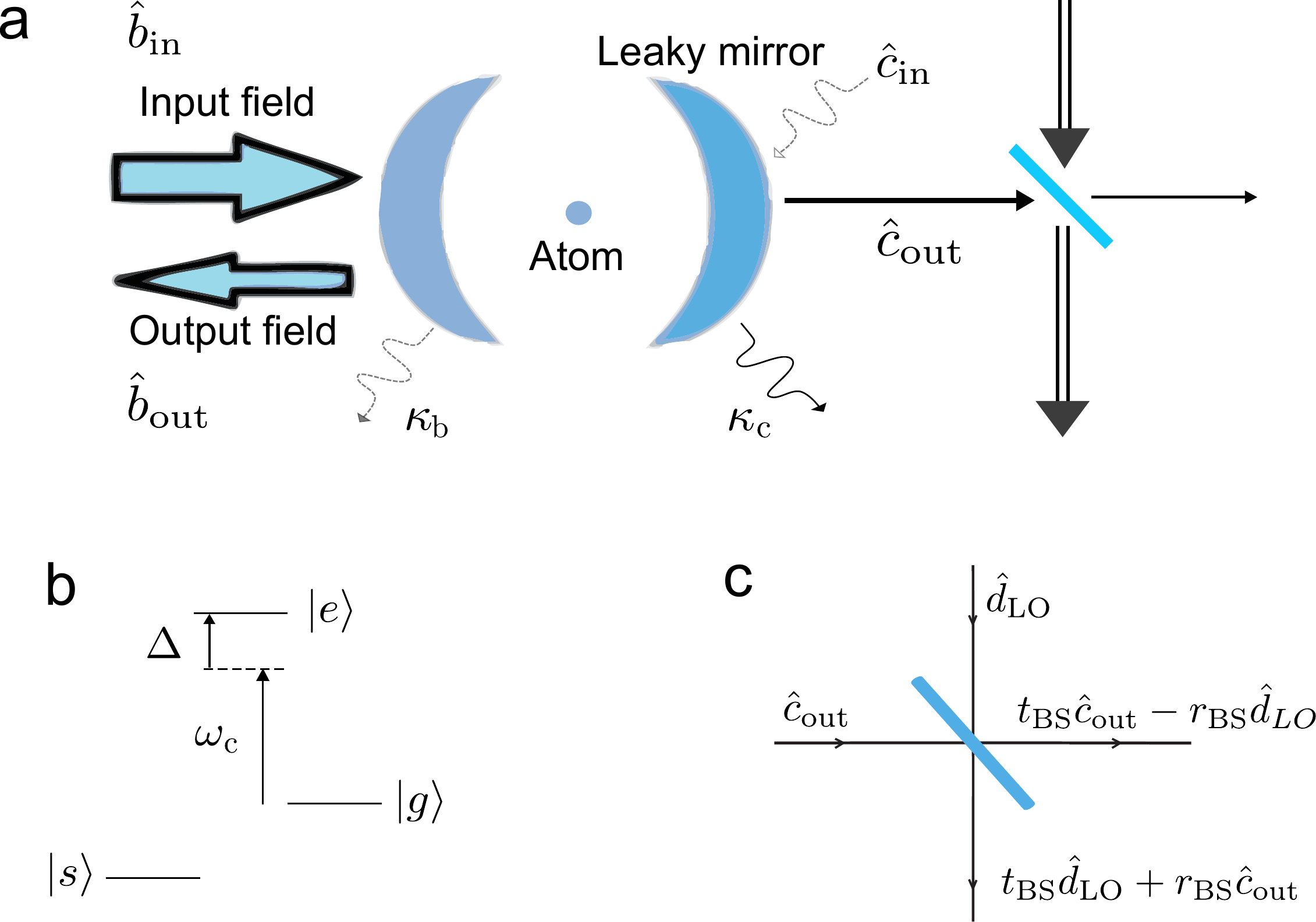}
\caption{
\textbf{State preparation.} \textbf{a.} An input field is incident on a cavity with an atom in the dispersive regime. This causes some reflection and some transmission of the cavity field. We assume that the mirror on the right has a lower reflectivity than the mirror on the left, such that the field predominantly leaves the cavity to the right. The final beam splitter performs a displacement operation which creates a superposition of propagating vacuum and coherent state. \textbf{b.} Level structure of the atom. The cavity field is dispersively coupled to the $\ket{g}-\ket{e}$ transition, and the $\ket{s}$ state is assumed to be far detuned such that it does not interact with the cavity field. \colin{{\bf c.} Beam splitter convention used. A displacement operation is applied on the mode labeled $\hat{c}_{\rm out}$ by combining it with a local oscillator (mode labeled $\hat{d}_{\rm LO}$). }} \label{fig:cavity}
\end{figure}

Since the cavity transmission depends on the atomic state, the atom-cavity system acts as a filter that transforms the initial state $\ket{\alpha_{\rm in}}\ket{\psi}_{\rm atom}$ in the following way:
\begin{align}
\ket{\alpha_{\rm in}} \otimes \ket{g} & \longrightarrow \ket{r_{\rm g} \alpha_{\rm in}}_{\rm refl} \otimes \ket{g} \otimes \ket{t_{\rm g} \alpha_{\rm in}}_{\rm trans} \\
\ket{\alpha_{\rm in}} \otimes \ket{s} & \longrightarrow \ket{r_{\rm s} \alpha_{\rm in}}_{\rm refl} \otimes \ket{s} \otimes \ket{t_{\rm s} \alpha_{\rm in}}_{\rm trans}
 \label{eqn:principle}
\end{align}
where $\ket{}_{\rm refl}$ and $\ket{}_{\rm trans}$ represent the reflected and transmitted photonic states, $\alpha_{\rm in}$ is the complex amplitude of the input coherent state and $r_j$ and $t_j$ ($j=s,g)$ are the amplitude reflectivity and transmitivity of the empty cavity (atom in the state $\ket{s}$, \ie $j=s$) and of the cavity with the atom (atom in the state $\ket{g}$,\ie $j=g$) and $|r_j|^2 + |t_j|^2 = 1$. Notice that we assume that both the reflected and transmitted fields are still coherent states (see Methods for details).

In order to produce a state of the form \eqref{eqn:target_state}, one needs to meet the following conditions: $t_{\rm s} = 0$ with $\braket{r_{\rm s} \alpha_{\rm in}}{r_{\rm g} \alpha_{\rm in}} \cong 1$ and $\ket{t_{\rm g} \alpha_{\rm in}} = \ket{\alpha}$, where $\ket{\alpha}$ is the desired coherent state in \eq{\ref{eqn:target_state}}. An alternative is to apply a displacement on the transmitted field \cite{matteo_g.a._displacement_1996}. \colin{The displacement is just the result of combining a field exiting the atom-cavity system ($\hat{c}_{\rm out}$ mode) with a coherent local oscillator $\ket{\beta}$ ($\hat{d}_{\rm LO}$ mode) on a beam splitter, as depicted in \fig{\subfigref[c]{fig:cavity}}}. The output port of interest is the port with the output field \colin{$\ket{t_{\rm BS}\hat{c}_{\rm out} - r_{\rm BS}\hat{d}_{\rm LO}}$}, where $r_{\rm BS}$ and $t_{\rm BS}$ are the amplitude reflectivity and transmitivity of the beam splitter respectively. \colin{For a coherent state, $\ket{\alpha_x}$, in the $\hat{c}_{\rm out}$ mode, one can achieve a zero amplitude in this output port by tuning the amplitude of $\ket{\beta}$ such that $r_{\rm BS}\beta = t_{\rm BS}\alpha_x$.} In our case, $\alpha_x=t_{\rm s}\alpha_{\rm in}$ and one can obtain the state $\ket{s,0}$ contained in \eq{\ref{eqn:target_state}}.

\begin{figure}[ht!]
\centering
\includegraphics[width=\columnwidth]{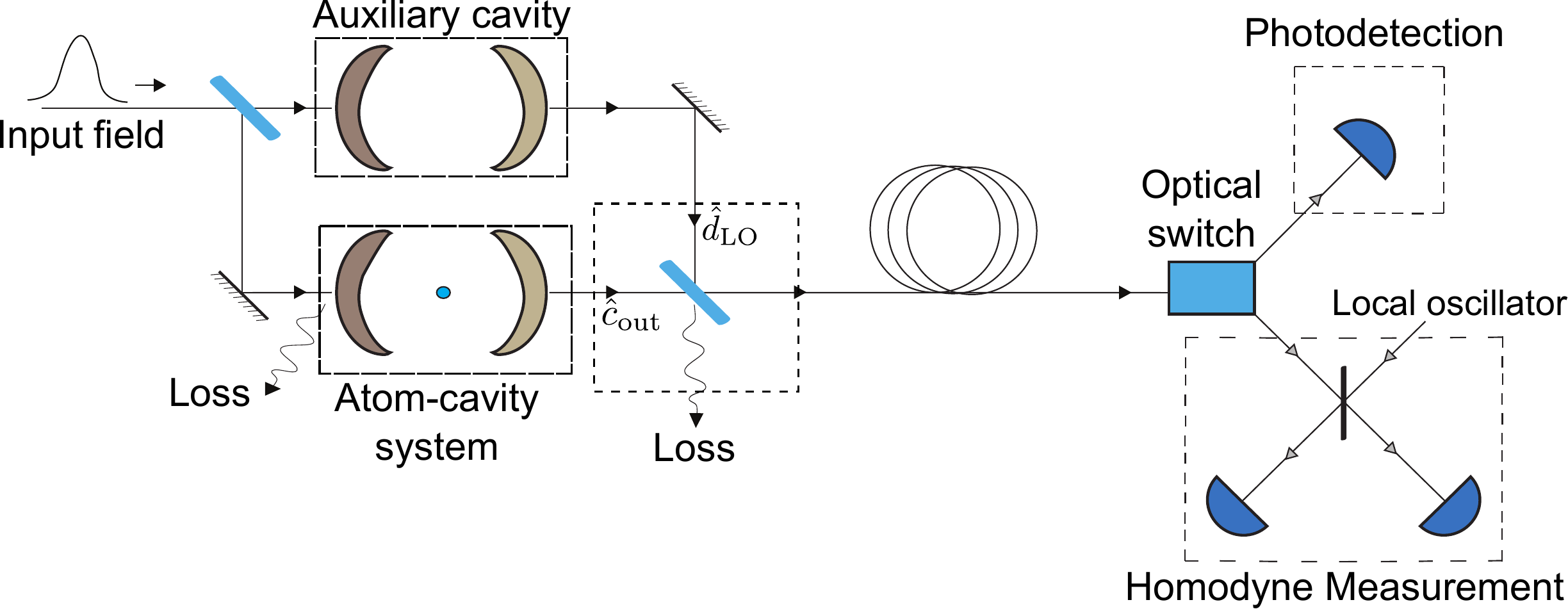}
\caption{
\textbf{Possible Experimental setup.} A pulse of light is first split on a beam splitter (blue). Part of it goes to the atom-cavity system (lower branch) described in Fig. \ref{fig:cavity}. The other part of the split pulse passes through an auxiliary cavity (upper branch) which is used to produce a pulse with the same spectral properties as that produced when the atom is in the $\ket{s}$ state. This requirement has to be met for a perfect displacement in the second beam splitter. Both beam splitters have the property that $|t_{\rm BS}|^2 \gg |r_{\rm BS}|^2$. After the state preparation, the photonic part of the state goes to a location that, given the times necessary for choosing the measurement and registering the results, assures space-like separation, and is submitted to either photodetection or homodyne measurement. Possible detection schemes of the atomic state are explained in the text (sec. \ref{sec:expt_feasible}).}
\label{fig:setup}
\end{figure}
The full Bell test setup is shown in \fig{\ref{fig:setup}}.
The beam splitters are identical with the property $|t_{\rm BS}|^2 \gg |r_{\rm BS}|^2$. Note that, in order to satisfy the condition $r_{\rm BS}\beta = t_{\rm BS}\alpha_x$, the spatiotemporal modes of the two coherent states $\alpha_x$ and $\beta$ must be identical. This can be achieved e.g. by using an auxiliary cavity similar to the one containing the atom (see Fig. \ref{fig:setup}), where $\alpha_x = t_{\rm s}\alpha_{\rm in}$. Eventually,  instead of using the auxiliary cavity, one might look for an atom-cavity-laser configuration in which the coupling of the field of polarization connecting the $\ket{s}$ state with some excited state is negligible compared to the dispersive coupling of the $\ket{g}-\ket{e}$ transition, and the orthogonal polarization coupling to neither $\ket{s}$ or $\ket{g}$ or $\ket{e}$. In such a situation one might use this field to produce the displacement pulse using only a single cavity.

The problem of an input coherent pulse with frequency spectrum $s_{\rm L}(\w)$ and amplitude $\alpha_{\rm in}$ impinging on a atom-cavity system can be treated using input-output theory \cite{walls_quantum_2008} (see Methods for details). This gives the form of the atomic state dependent amplitude reflection and transmission coefficients, $r_j(\w)$ and $t_j(\w)$, $j=s,g$ as functions of the input pulse.

\dani{The final atom-field state is obtained upon tracing away non-radiative losses in the cavity mirrors, the fields reflected by the cavity, emitted by the atom and the other output port of the beam splitter. This naturally leads to coherence (and entanglement) loss and results, when the atom is initially prepared in the state $\cos\nu \ket{s} + \sin\nu e^{i \phi}\ket{g}$,  in the mixed state}
\begin{equation}
  \rho = V \proj{\psi_{\rm f}}{\psi_{\rm f}} + (1-V) \sigma, \label{eqn:real_state}
\end{equation}
where
\begin{align}
   \ket{\psi_{\rm f}} &= \cos\nu \ket{s,0} + \sin\nu  \ket{g,\{\tilde{\alpha}\}}, \label{eqn:final_state}\\
   \sigma &= \cos^2\nu \proj{s,0}{s,0} + \sin^2\nu \proj{g,\{\tilde{\alpha}\}}{g,\{\tilde{\alpha}\}},
\end{align}
and the visibility $V$ is (see Methods for details),
\begin{align}
  V &= \exp\Big[-F \frac{|\tilde{\alpha}|^2}{2 t_{\rm BS}^2} \Big],  \label{eqn:vis}\\
  F&= r_{\rm BS}^2+f_{\rm cav} + I_{s_{\rm L}} \inv{4 C} (1+f_{\rm cav}), \label{eqn:vF} \\
  I_{s_{\rm L}} &= \frac{\int d\w |s_{\rm L}(\w)\inv{D(\w)}|^2}{\int d\w |s_{\rm L}(\w)\inv{D(\w)} \inv{1+i2 (\w-\w_{\rm c})/\kappa}|^2}, \\
  D(\w) &=(\half\Gamma + i(\w-\w_{\rm ge}))(\kappa /2 +i(\w-\w_{\rm c})) +g^2, \label{eqn:D_w}
\end{align}
where $C$ is the usual single-atom cooperativity $C=\frac{g^2}{\Gamma \kappa}$ and $f_{\rm cav} = \frac{\kappa_{\rm b} + \kappa_{\rm L}}{\kappa_{\rm c}} $ is a factor which describes the asymmetry of the cavity. Here, the continuous frequency coherent state, $\ket{\{\tilde{\alpha}\}}$, is
\begin{align}
  \ket{\{\tilde{\alpha}\}} &= {\rm exp} \big[ t_{\rm BS}\alpha_{\rm in} \int d\w\, s_{\rm L}(\w) \pare{t_{\rm g}(\w) - t_{\rm s}(\w)} c\dagg_{\rm out}(\w)  +h.c.\big] \ket{0}, \\
  |\tilde{\alpha}|^2 &= |t_{\rm BS}\alpha_{\rm in}|^2 \int d\w \, \big| s_{\rm L}(\w) (t_{\rm g}(\w) - t_{\rm s}(\w))\big|^2, \label{eqn:alpha2}
\end{align}
with the transmission coefficients
\begin{align}
  t_{\rm s}(\w) &= \frac{\sqrt{\kappa_{\rm b} \kappa_{\rm c}}}{\kappa /2+i(\w-\w_{\rm c})}, \label{eqn:ts}\\
  t_{\rm g}(\w) &= \frac{\sqrt{\kappa_{\rm b}\kappa_{\rm c}}}{D(\w)}\pare{\half\Gamma + i(\w-\w_{\rm ge})}, \label{eqn:tg}
\end{align}
and the atomic state is initially prepared with $\phi$ such that the final state is of the form \eqref{eqn:final_state}. In the above, $s_{\rm L}(\w)$ is the frequency spectrum of the laser field, $g$ is the coupling constant of the cavity mode to the $\ket{g}-\ket{e}$ transition, $\Gamma$ is the transverse decay rate of the $\ket{e}$ state and $\kappa_{\rm L}$ is the loss rate of the cavity mirrors.

From equation \eqref{eqn:vis}, it is easy to see that for $F\to 0$, we have $V \to 1$. Also for $F\to0$, we need the 3 conditions, $r_{\rm BS} \to 0$, $f_{\rm cav}\to 0$ and $C\to \infty$. In practice, the first condition means that one should use a small value of the beam splitter reflectivity and adjust the amplitude of the local oscillator, such that the condition $r_{\rm BS} \beta = t_{\rm BS} t_{\rm s} \alpha_{\rm in}$ is still satisfied. The second condition is precisely the requirement of the asymmetric cavity we have mentioned at the beginning of this section and is dependent only on the transmission and losses of the cavity mirrors used. For completeness, we have also included non-radiative mirror losses. The third condition means that one needs large single-atom cooperativity. One remarkable feature of this result is that for $C\gg 1$, the visibility $V$ (but of course not the state) is independent of the details of the spectrum of the input pulse.

The final point to note is that to satisfy the approximations used in our derivation, we require a negligible probability of exciting the atom throughout the duration of the input pulse. This is important, since, to maintain some specific amplitude of the output coherent state $\ket{\{\tilde{\alpha}\}}$ after the beam splitter, one might have to use a very large amplitude input laser if the integral in equation \eqref{eqn:alpha2} is small. For the sake of this theoretical analysis, we consider the case of an input Gaussian pulse with the normalized spectrum
\begin{equation}
  |s_{\rm L}(\w)|^2 = \inv{\gamma_{\rm L}\sqrt{\pi}} e^{-\pare{\frac{\w-\w_{\rm L}}{\gamma_{\rm L}}}^2},
\end{equation}
where $\w_{\rm L}$ is the central laser frequency, and $\gamma_{\rm L}$ is the bandwidth of the pulse. \comm{The spectral shape of the input pulse we have chosen is arbitrary and the results should depend only on the pulse bandwidth.}

To close the locality loophole, one has to propagate the outgoing pulse for some minimum distance, given by the larger of the measurement times of the atom and the pulse, which depends on the pulse duration. The pulse duration is given by the relevant time scale of the system, \ie, either by the input pulse duration or the cavity lifetime, whichever is greater. This means that, from the point of view of closing the locality loophole, it is useful to push the duration of the input pulse down to the cavity lifetime, but not necessarily further. Since the pulse duration and the cavity lifetime are proportional to the inverse of the pulse bandwidth $\gamma_{\rm L}$ and the cavity decay rate $\kappa$ respectively, the best case would be if $\gamma_{\rm L} \approx \kappa$.

\begin{figure}[ht!]
\centering
\includegraphics[width = \columnwidth]{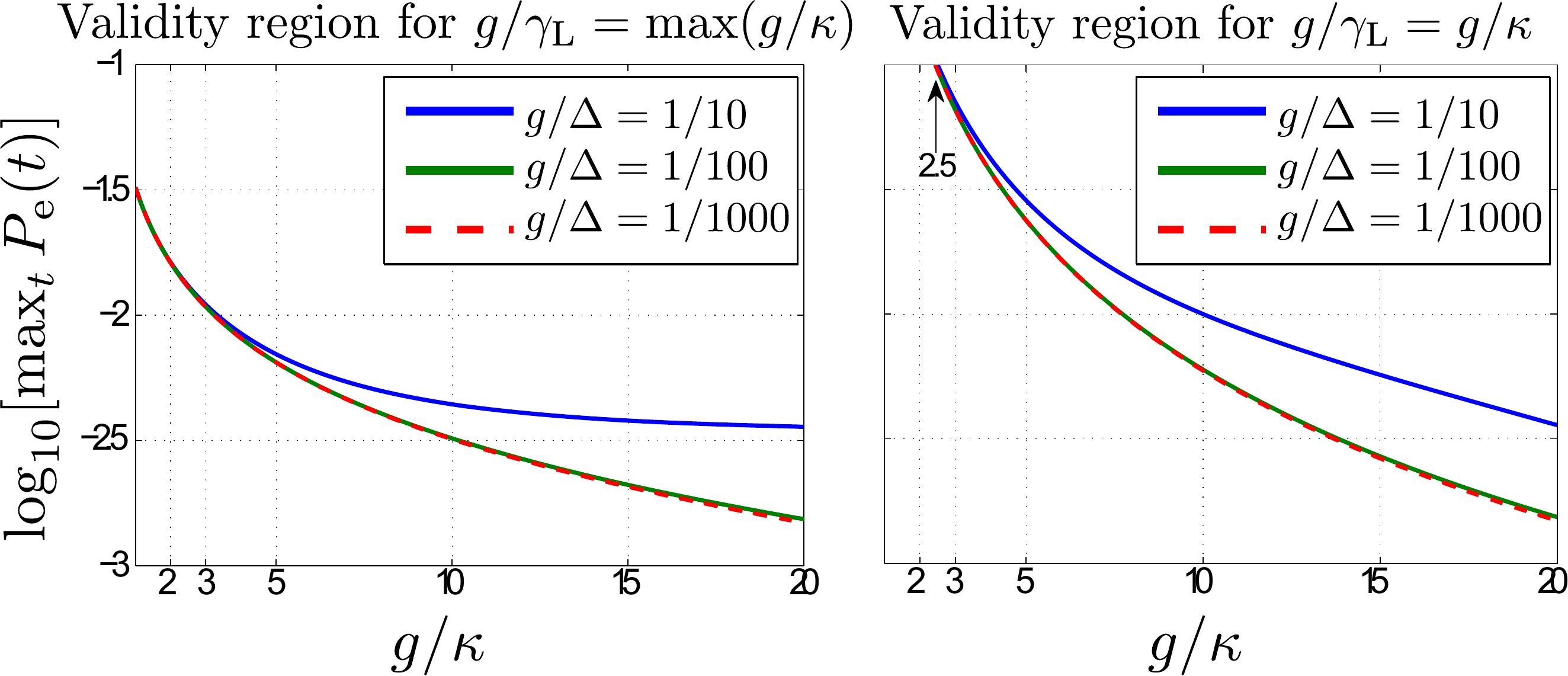}
\caption{
\textbf{Plots of the maximum probability of exciting the atom over the duration of the input pulse.} This figure shows plots of $\log_{10}[\max_t P_{\rm e}(t)]$ as a function of the parameter $\frac{g}{\kappa}$, assuming that the central laser frequency is on resonance with the bare cavity, for different values of $\frac{g}{\Delta}$. The left panel is the case with input pulse with minimal bandwidth setting $g/\gamma_{\rm L} = \max(g/\kappa)$. The right panel is the case with the input pulse bandwidth set equal to the cavity bandwidth ($g/\gamma_{\rm L} = g/\kappa$). \colin{For $g/\Delta$ small enough, the probability of exciting the atom becomes insensitive to the precise value of $g/\Delta$. This fact is illustrated by the overlap of curves corresponding to $g/\Delta=1/100$ and $g/\Delta=1/1000$.} The parameters used in this plot are $|\alpha|=2.1$, $f_{\rm cav} = 4/100$ and $g/\Gamma=5/3$.} \label{fig:validity}
\end{figure}
Fig.~\ref{fig:validity} shows plots of $\log_{10} (\max_t P_{\rm e}(t))$, the maximum probability of exciting the atom
as a function of the parameter $\frac{g}{\kappa}$ (the coupling strength over the total cavity decay rate), assuming that the central laser frequency is on resonance with the bare cavity, for $\frac{g}{\Delta}=1/10,1/100,1/1000$. The left figure is the case for a fixed minimal bandwidth of the input pulse in the sense that we choose $\gamma_{\rm L}$ such that $g/\gamma_{\rm L} = \max(g/\kappa)$. The right figure corresponds to the situation when the input pulse bandwidth is set equal to the cavity bandwidth in order to minimize the pulse duration. In the following, we will consider the regions where $(\max_t P_{\rm e}(t))\leq 0.1$, to be the parameter regimes when our analysis is suitable.
 As is evident from Fig.~\ref{fig:validity}, for some large $g/\gamma_{\rm L}$, there always exists some $g/\kappa$ for which our solutions are self-consistent, which is not the case when $\gamma_{\rm L} = \kappa$, the situation minimizing the pulse duration. We can thus look for some intermediate value of $\gamma_{\rm L}$ (having all the other parameters fixed), which minimizes the pulse duration while still respecting the validity of the used approximation. This is the strategy we employ in the next section, where we will discuss a possible implementation of our scheme using realistic experimental parameters.

\subsection*{Experimental feasibility}  \label{sec:expt_feasible}
So far, we theoretically described a general cavity scenario. In the following, we will focus our attention on possible implementations of our scheme in optical cavities. This implementation has the advantage of allowing large propagation distances due to the availability of low loss optical fibers.

We start by general constraints on measurement times and propagation distances imposed by the locality loophole. To close the locality loophole, one requires the start of the measurement event on one side to be space-like separated {\col{from}} the end of the measurement event on the other side. In other words, we need that the space-time coordinates of the end of the atomic measurement and the choice of measurement basis (photon detection or homodyning) on the photonic side be space-like separated, and vice versa. This constraint translates into the minimum {\col{propagation}} distance
\begin{equation}
  d \geq c\pare{\max\pare{(\Delta t_{\rm ph,c} + \Delta t_{\rm ph,m}) , (\Delta t_{\rm at,c} + \Delta t_{\rm at,m})}}, \label{eqn:locality}
\end{equation}
where $c$ is the speed of light in free-space, $\Delta t_{\rm at/ph,c}$ is the time required on the atomic/photonic side to choose the measurement settings, which also includes the time to change the measurement basis, $\Delta t_{\rm at/ph,m}$ is the time required on the atomic/photonic side to perform the actual measurement, which necessarily includes the duration of the light pulse. We assume that the main time constraint is set by the atomic measurement, which is typically slower than the measurement of photons. The above equation shows, that even if we used a very short pulse, we would still need to propagate it at least to a distance, given by the sum of atomic measurement time and the time required to choose a basis (a rotation in the space of $\ket{s}$ and $\ket{g}$, {\col{typically using RF pulses}}), multiplied by $c$. The best trade-off in this case, is thus to have $\Delta t_{\rm ph,m} \approx \Delta t_{\rm at,m}+\Delta t_{\rm at,c}$. This is possible because the time needed to choose a basis on the photonic side can be very fast \cite{weihs_violation_1998}.

One possibility to perform a fast atomic state detection would be to use a 2-photon ionization technique \cite{henkel2010highly}. This technique has been performed in free-space configurations and can achieve 98\% efficiency in < \SI{1}{\micro s}. Although the restriction of a small cavity might make such a photo-ionization technique technically challenging, proposals to make focusing cavities with large physical volumes (lengths of $\sim$ 1\,cm) while still maintaining large coupling constants ($g \sim$ 100\,MHz) exist \cite{syed2010,morrin1994} potentially making such techniques compatible. Here, we assume $\Delta t_{\rm at,m}+\Delta t_{\rm at,c} = \SI{1}{\micro s}$. We should thus seek to produce a pulse which requires a total measurement duration up to \SI{1}{\micro s}. 
{\col{For the Gaussian pulse example considered in our calculation, we take the total pulse measurement duration to be $\Delta t_{\rm ph,m} = 6\sqrt{2} / \gamma_{\rm L,m}$. The factor $6\sqrt{2}$ is chosen such that if the bandwidth of the pulse incident on the cavity is $\gamma_{\rm L,m}$, then the neglected tails of the outgoing pulse correspond to <0.5\% of the total integrated intensity of the outgoing pulse (for the parameters used in Table \ref{tab:parameters}, see below). For a measurement time of $\SI{1}{\micro s}$, this corresponds to $\gamma_{\rm L,m}= 2\pi \times 1.35\textrm{MHz}$. It also means that for outgoing pulse bandwidths $\gamma_{\rm L} > \gamma_{\rm L,m}$ (shorter pulses), the total measurement time is set by $\Delta t_{\rm at,m} + \Delta t_{\rm at,c} = \SI{1}{\micro s}$ and the minimum propagation distance is 300 m.}}

Next, we use experimental parameters given in \cite{birnbaum2005photon,hood_thesis,ritter_elementary_2012}. These experiments use \SI{852}{nm} and \SI{780}{nm} light respectively, which are subject to about \SI{2}{dB/km} loss in optical fibers. In the following, we compare both experiments and possible modifications. As discussed previously, we will work in the large detuning regime, $\Delta \gg g$. We choose an arbitrary, but subjected to experimental constraints, value of $g/\Delta = 1/10$. This ensures that we respect all approximations used, as long as we also satisfy condition $\max_t P_{\rm e}(t) \leq 0.1$. {\col{The figure of merit in both cases will be $\gamma_{0.1}$, which is the largest acceptable laser bandwidth that satisfies condition $\max_t P_{\rm e}(t) \approx 0.1$, and the state production visibility \eqref{eqn:vis}, which is computed using the actual laser bandwidth $\equiv \min(\kappa,\gamma_{0.1},\gamma_{\rm L,m})$. We also include the effect of finite pulse measurement time on the visibility, using \eqref{eqn:vis}-\eqref{eqn:D_w}. We assume $|\alpha|=2.1$ and $|r_{\rm BS}|^2 =0.001$. Table \ref{tab:parameters} summarizes the results.}}

\begin{table} [h!]
\begin{center}
\begin{tabular}{*{7}{|c}|}
\hline
  $g/(2\pi)$ & $\kappa/(2\pi)$  & $\Gamma/(2\pi)$  & $f_{\rm cav}$ & $\gamma_{0.1}/(2\pi)$ & $V$ &$d_{min}$\\
  \hline
    34 MHz & 4.1 MHz & 2.6 MHz & 10/4 & 13.3 MHz & 0.4\% & 300 m\\
   \hline
   34 MHz & 4.1 MHz & 2.6 MHz & 14/100 & 63.3 MHz & 72.7\%  & 300 m \\
   \hline
   34 MHz & 4.1 MHz & 2.6 MHz & 4/100 & 65.2 MHz & 90.8\%& 300 m \\
   \hline
   5 MHz & 3 MHz & 3 MHz & 14/100    & 1.1 MHz & 56.3\%& 370 m\\
   \hline
   5 MHz & 3 MHz & 3 MHz & 4/100    & 1.3 MHz & 71.3\%& 310 m\\
   \hline
   5 MHz & 1.5 MHz & 3 MHz & 4/100 & 3.1 MHz & 77.2\% & 300 m\\
   \hline
\end{tabular} \caption{\textbf{Expected visibilities {\col{and minimum propagation distances}} for available experimental parameters for $^{133}$Cs (first 3 rows) and $^{87}$Rb (last 3 rows).} All parameters ($g/(2\pi),\kappa/(2\pi),\Gamma/(2\pi),f_{\rm cav}$) in the first and fourth row are actual cavity parameters (including mirror losses) obtained from \cite{birnbaum2005photon,hood_thesis,ritter_elementary_2012}. The second row shows the effect on the visibility and $\gamma_{0.1}$ by reducing $f_{\rm cav}$ to the current value in the experiment of Ref.~\cite{ritter_elementary_2012}. The third row is obtained by neglecting the mirror losses, which further decreases the value of $f_{\rm cav}$. The fifth row shows the effect of neglecting mirror losses, and the last row shows the effect of increasing $g/\kappa$. {\col{Notice that $\gamma_{0.1}>2\pi \times \SI{1.35}{MHz}$ in the first three and last rows, which means that the propagation distance is limited by the detection time on the atomic side and not by the pulse duration and corresponds to the propagation distance of 300 m.}} The state production visibility $V$ is computed assuming $|\alpha|=2.1$, $|r_{\rm BS}|^2 = 0.001$, using the laser bandwidth $\min(\kappa,\gamma_{0.1},\gamma_{\rm L,m})$ \colin{and taking into account the truncation of the pulse due to finite measurement time.} 
}
\label{tab:parameters}
\end{center}
\end{table}
The parameters of Ref.~\cite{birnbaum2005photon,hood_thesis} are $(g/(2\pi),\kappa/(2\pi),\Gamma/(2\pi),f_{\rm cav})$ = (34 MHz, $\SI{4.1}{MHz}$, 2.6\,MHz, 10/4) (we assume that the experiment performed implements a symmetric cavity). Due to the symmetry of the cavity, one obtains a small effective visibility. Assuming that the cavity could be made asymmetric reducing $f_{\rm cav}$ to the current value in the experiment of Ref.~\cite{ritter_elementary_2012}, the visibility dramatically increases to 72.7\% (second row). Neglecting the mirror losses, we show in the third row that $V$ can be as high as 90.8\%.

The parameters of Ref.~\cite{ritter_elementary_2012} are $(g/(2\pi),\kappa/(2\pi),\Gamma/(2\pi),f_{\rm cav})$ = ($\SI{5}{MHz}$, $\SI{3}{MHz}$, $\SI{3}{MHz}$, 14/100). As the setup stands, the visibility is 56.3\%. However, neglecting the mirror losses, $V$ increases to 71.3\% (fifth row). If it were further possible to reduce the total cavity decay rate by a factor of 2, thus increasing the cooperativity, while maintaining the same asymmetry, the visibility further increases to 77.2\% (sixth row). The required incident photon number can be calculated from \eqref{eqn:alpha2}. For the parameters in Table \ref{tab:parameters} and requiring the resulting photon number, $|\tilde{\alpha}|^2 = 2.1^2$, one requires $|\alpha_{\rm in}|^2\approx 25-400 $ input photons.

For the specific case of $^{87}$Rb, one might also identify possible states playing the role of $\ket{g}$, $\ket{s}$ and $\ket{e}$. We may choose for example the $\ket{s}$ state to be the hyperfine state $\ket{5S_{1/2}, F = 1, m_F = 1}$, the $\ket{g}$ state $\ket{5S_{1/2}, F = 2, m_F = 2}$ and the $\ket{e}$ state $\ket{5P_{3/2}, F = 3, m_F = 3}$. In this case, the input pulse and cavity field would have a $\sig_+$ polarization coupling the $\ket{g}-\ket{e}$ transition. Due to the large detuning of the hyperfine states (\SI{6.8}{GHz}), and the fact that the $s$ state is far-detuned to any other $\sig_+$ transitions, these states are possible candidates for the experiment.
\begin{figure}[ht]
\centering
	\begin{tikzpicture}
	\begin{axis}[xlabel=$\eta$,ylabel=$|t_{\rm line}|^2$,legend style={anchor=north west,cells={anchor=west},at={(0.05,.5)}}]
	\addplot[color=magenta,mark=none] coordinates
{(1,0.67701)(0.961,0.69014)(0.92206,0.705)(0.884,0.71903)(0.846,0.73405)(0.809,0.75)(0.772,0.766)(0.735,0.78211)(0.699,0.79912)(0.664,0.817)(0.629,0.835)
(0.59401,0.854)(0.56003,0.873)(0.527,0.89207)(0.49402,0.913)(0.46201,0.934)(0.431,0.955)(0.4,0.977)(0.36901,1)};
	\addplot[color=green,mark=none] coordinates
{(1,0.82699)(0.95,0.84015)(0.899,0.85456)(0.849,0.86973)(0.8,0.88567)(0.752,0.90241)(0.70489,0.92)(0.658,0.93875)(0.613,0.958)(0.568,0.9786)(0.52431,1)};
	\addplot[color=red,mark=none] coordinates
{(1,0.91154)(0.949,0.92114)(0.89719,0.932)(0.847,0.94366)(0.798,0.9562)(0.749,0.96998)(0.702,0.98445)(0.65568,1)};
	\addplot[color=black,mark=none] coordinates
	 {(1,0.98057)(0.97312,0.984)(0.947,0.9876)(0.921,0.99145)(0.895,0.99558)(0.86903,1)};
	\addplot[color=blue,mark=none,dashed] coordinates
{(1,0.53701)(0.99631,0.538)(0.91466,0.561)(0.834,0.58592)(0.756,0.61234)(0.67932,0.641)(0.605,0.67133)(0.53292,0.704)
(0.46235,0.739)(0.395,0.776)(0.329,0.81544)(0.266,0.858)(0.206,0.90201)(0.148,0.94901)(0.093,0.999)(0.092001,1)};%

\node[color=magenta,font=\small] at (axis cs:0.37, 1.02) {2};
\node[color=green,font=\small] at (axis cs:0.53, 1.02) {2.05};
\node[color=red,font=\small] at (axis cs:0.655, 1.02) {2.1};
\node[color=black,font=\small] at (axis cs:0.87, 1.02) {2.15};
\node[color=blue,font=\small] at (axis cs:0.092,1.02) {$\ket{\psi_\alpha}$};

	\addplot[only marks,color=black,mark=x] coordinates
	{
	(0.629,0.835) (0.752,0.90241) (1, 1)};

\node[color=black,font=\small] at (axis cs:0.97, 1) {A};
\node[color=black,font=\small] at (axis cs:0.7, 0.90241) {B};
\node[color=black,font=\small] at (axis cs:0.58, 0.835) {C};
	\end{axis}
	\end{tikzpicture}

\caption{
\textbf{Contour lines of $\mean{\chsh}$ as a function of $\eta$ and $|t_{\rm line}|^2$.} In this plot we fixed $|\alpha|=2.1$  and optimized the measurement and state parameters $\gamma, b, \nu$ for each point. The parameters used are $(g/(2\pi),\kappa/(2\pi),\Gamma/(2\pi),f_{\rm cav})$ = (34\,MHz, 4.1\,MHz, 2.6\,MHz, 14/100) and $\abs{r_{\rm BS}}^2 = 0.001$ (second row of Table \ref{tab:parameters}). The result for the ideal state in \eqref{eqn:target_state} (dotted line) is included for comparison. \colin{For the sake of illustration we give the specific numbers for the points represented by crosses A, B and C in Table \ref{tab:realistic_pts}.}
} \label{contourlines}
\end{figure}
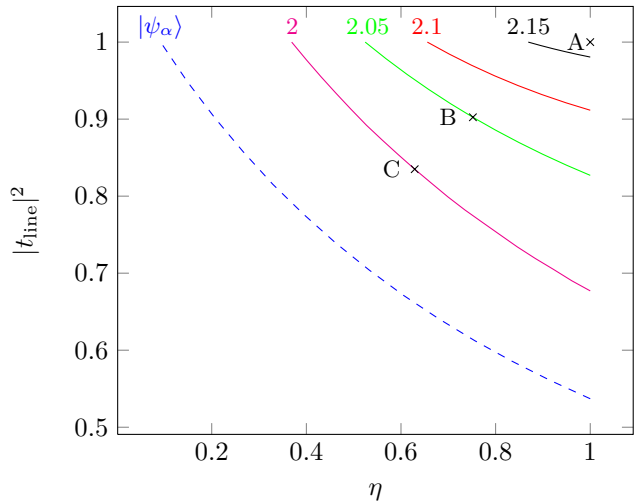

\begin{table}[h!]
\centering
\begin{tabular}[b]{*{4}{|c}|}
\hline
 & A & B & C\\
 \hline
  $|t_{\rm line}|^2$ & 1 & 0.9 & 0.83 \\
 \hline
  $\eta$ &1 & 0.75 & 0.63\\
  \hline
 $\gamma$ & 0.42 & 0.33 & 0.02 \\
 \hline
 $\nu$ & 0.75 & 0.6 & 0.03 \\
 \hline
 $b$ & 0.53 & 0.56 & 0.58 \\
 \hline
  $\mean{\chsh}$ & 2.17 & 2.05 & $2^+$\\
   \hline
\end{tabular}
\caption{{\bf Optimized parameters for 3 points on \fig{\ref{contourlines}}.} These parameters are results from numerical optimization of the state \eqref{eqn:real_state}, given some detector efficiency $\eta$ and some transmission $|t_{\rm line}|^2$. We have fixed $|\alpha|=2.1$, and used the Visibility $V$ \eqref{eqn:vis} from the second row of Table \ref{tab:parameters}.} \label{tab:realistic_pts}
\end{table}
We now maximize $\mean{\chsh}$ as a function of transmission and detector inefficiency for a set of realistic parameters. We thus set $|\alpha| = 2.1$,  $(g/(2\pi),\kappa/(2\pi),\Gamma/(2\pi),f_{\rm cav})$ = (34\,MHz, 4.1\,MHz, 2.6\,MHz, 14/100) and $\abs{r_{\rm BS}}^2 = 0.001$ (second row of Table \ref{tab:parameters}). Notice that we have included possible non-radiative losses from cavity mirrors.  Given these constraints, the maximal CHSH violation
is $2.17$. Moreover, we can attain a violation even for
$\eta=0.37$ with $|t_{\rm line}|^2=1$, or transmittance
$|t_{\rm line}|^2=0.68$ with $\eta=1$.
\fig{\ref{contourlines}} summarizes these results. Note that the higher the visibility of the produced state, the closer one gets to the ideal scenario (dashed line in \fig{\ref{contourlines}})

\section*{Discussion}

In this work, we have shown that using current technology, it is possible to produce a hybrid atom-photon entangled state that still violates the CHSH inequality up to a value of 2.17. We also showed that we can attain a violation even for a low photon counting efficiency of 37\% with perfect transmission, or a line transmission of 68\% with perfect detection efficiency. Moreover, assuming that it is possible to perform photoionization measurements < \SI{1}{\micro s}, the required propagation distances to close the locality loophole are of the order of 300m for optical setups. This gives a very good outlook for optical systems in eventually performing a loophole-free Bell test.

In principle, one may also consider circuit QED setups. The advantages of such setups are twofold. First, the very fast and efficient qubit state detection, on the order of \SI{10}{ns} \cite{ansmann_violation_2009} potentially lowers the required propagation distances to laboratory scale distances (\SI{10}{m} or less). Second, in this system, large ratios of coupling constant to cavity decay, $g/\kappa$, can be achieved. The drawbacks with current technology are the limited efficiencies of both photodetection and homodyne detection (private communication with Steve Girvin), 
 and the requirement to cool the propagation line down to cryogenic temperatures. The hope is that both these drawbacks, being of technological nature, can be eventually overcome in near future experiments.

Before finishing let us compare our present results with previous proposals involving similar setups. In Ref. \cite{sangouard11} a Bell test involving the production of an entangled state between an atom and a photonic field created by atomic decay was studied (see also \cite{BC12}). It was shown that Bell violations with photodetection efficiency of $39\%$ (see green curve in Fig.~\ref{comparison}) can be achieved.  Note however that this number refers to the ideal state, considering that all light emitted by the atom is collected. Our scheme overcomes this problem since the photonic part of the state is the transmitted mode, and, moreover, requires lower efficiencies.

We believe that our proposal will trigger possible implementations of loophole-free Bell tests with atom-photon interfaces, which are particularly important in quantum communication and cryptographic applications.
\section*{Methods}
\subsection*{Input-output relations}
Here we derive the input-output relations \cite{walls_quantum_2008} for a two-sided cavity, clearly stating the approximations used, and their validity.
We consider a cavity mode ($a$ mode) which couples to a left mode ($b$ mode) and a right mode ($c$ mode). We also include loss in the mirrors of the cavity as the coupling of the cavity $a$ mode to an additional $L$ mode.
\\
These give the following Heisenberg equation of motion for the cavity field,
\begin{align}
  \partial_t a(t) &= -\frac{i}{\hbar} \com{H_{\rm sys}}{a(t)} - \frac{\kappa}{2} a(t) {} \nn\\
  & {} + \sqrt{\kappa_{\rm b}}b_{\rm in}(t) + \sqrt{\kappa_{\rm c}}c_{\rm in}(t) + \sqrt{\kappa_{\rm L}}L_{\rm in}(t), \label{aeqn:starting}
\end{align}
where $\kappa = \kappa_{\rm b} +\kappa_{\rm c} + \kappa_{\rm L}$ and $H_{\rm sys}$ is the system Hamiltonian (empty cavity or cavity with atom) without considering the baths, which is either
\begin{align}
  H_{\rm empty} &= \hbar \w_{\rm c} a\dagg a \quad \textrm{or,}\\
  H_{\rm atom-cavity} &= \hbar \w_{\rm c} a\dagg a + \hbar \w_{\rm a} \sig\dagg \sig + \hbar g (a\dagg \sig + a \sig\dagg).
\end{align}
Equation \eqref{aeqn:starting} can also be written in the equivalent way
\begin{align}
  \partial_t a(t) &= -\frac{i}{\hbar} \com{H_{\rm sys}}{a(t)} - \frac{k_{\rm c}}{2} a(t) {} \nn \\
  & {}+ \sqrt{\kappa_{\rm b}}b_{\rm in}(t) - \sqrt{\kappa_{\rm c}}c_{\rm out}(t) + \sqrt{\kappa_{\rm L}}L_{\rm in}(t),
\end{align}
where $k_\alpha = \kappa-2\kappa_\alpha$ and we now consider the output operator for the $c$ mode, where the input and output operators are defined as,
\begin{align}
  O_{\rm in}(t) &= \frac{-1}{\sqrt{2\pi}} \int d\w \, O_\w (t_0) e^{-i\w(t-t_0)}, \\
  O_{\rm out}(t) &=   \inv{\sqrt{2\pi}} \int d\w \, O_\w (t_1) e^{-i\w(t-t_1)}.
\end{align}
For the case of an empty cavity, these equations are just a linear system of differential equations, and can be solved simply by Fourier transforms defined as
\begin{align}
  f(t) &= \inv{\sqrt{2\pi}} \int \dint\w\,e^{-i\w t}f(\w), \\
  f(\w) &= \inv{\sqrt{2\pi}} \int \dint t \,e^{i\w t}f(t),
\end{align} to give
\begin{eqnarray}
\begin{pmatrix}
  b_{\rm in}(\w) \\
  c_{\rm in}(\w) \\
  L_{\rm in}(\w)
\end{pmatrix} =
{\cal U}_{\rm s} \begin{pmatrix}
  b_{\rm out}(\w)\\
  c_{\rm out}(\w) \\
  L_{\rm out}(\w)
\end{pmatrix}, \label{aeqn:emptyio}
\end{eqnarray}
where
\begin{align}
  {\cal U}_{\rm s} &= \begin{pmatrix}
    r_{\rm b}(\w) & t_{\rm b}(\w) & l_{\rm b}(\w) \\
    t_{\rm b}(\w) & r_{\rm c}(\w) & l_{\rm c}(\w)\\
    l_{\rm b}(\w) & l_{\rm c}(\w) & r_{\rm L}(\w)
  \end{pmatrix} \label{aeqn:cavU} \\
  r_\alpha(\w) &= - \big(\frac{\frac{k_\alpha}{2}+ i (\w-\w_{\rm c})}{\frac{\kappa}{2}+i(\w-\w_{\rm c})}\big) \\
  t_{\rm b}(\w) &= \frac{\sqrt{\kappa_{\rm b}\kappa_{\rm c}}}{\frac{\kappa}{2}+i(\w-\w_{\rm c})} \\
  l_{\rm b/c}(\w) &= \frac{\sqrt{\kappa_{\rm L}\kappa_{\rm b/c}}}{\frac{\kappa}{2}+i(\w-\w_{\rm c})}
\end{align}
 These are relations used in the main text when the atom is in the $\ket{s}$ state, since it is assumed to be decoupled from all cavity and environmental modes. However, when the atom is in the $\ket{g}$ state, it is assumed to couple to the cavity $a$ mode. In this situation, one has to include atomic spontaneous emission. This can be done by including the coupling of the $\ket{g}-\ket{e}$ transition to an additional bath, $E$ mode, different from the cavity. This gives the set of equations
\begin{align}
  \partial_t a(t) &= -\frac{i}{\hbar} \com{H_{\rm atom-cavity}}{a(t)} - \frac{\kappa}{2} a(t) {} \nn \\
  & {} + \sqrt{\kappa_{\rm b}}b_{\rm in}(t) + \sqrt{\kappa_{\rm c}}c_{\rm in}(t) + \sqrt{\kappa_{\rm L}}L_{\rm in}(t), \\
  \partial_t \sig(t) &= -\frac{i}{\hbar} \com{H_{\rm atom-cavity}}{\sig(t)} - \frac{\Gamma}{2} \sig(t) - \sqrt{\Gamma}\sig_z E_{\rm in}(t),
\end{align}
where $\Gamma$ describes the rate of emission into modes other than the cavity modes, and $\sig_{\rm in}$ is the input operator of the environment. Note that $\Gamma$ can be made small if the physical cavity mode has a large spatial overlap with the emission pattern of the $\ket{g}-\ket{e}$ transition.

In the following, we will investigate the system dynamics in the low excitation regime. The reason is twofold. First, we want to avoid exciting the atom in order to prevent the decay into the environment (spontaneous emission with the decay rate $\Gamma$). Second, populating the excited state would induce a more complex dynamics producing in general a nontrivial entangled state between the atom and input, output and cavity fields. This is certainly an interesting regime to investigate in the context of quantum state engineering, but in the present paper we focus on the more intuitive picture in the spirit of \eq{\ref{eqn:principle}}. The assumption of the atom occupying mostly the ground state can be translated as $\sig_z \approx -\one$. With this assumption, the above set of equations can be solved analytically. Proceeding analogously to the empty cavity case, we have
\begin{equation}
  \begin{pmatrix}
    b_{\rm in}(\w) \\
    c_{\rm in}(\w) \\
    L_{\rm in}(\w) \\
    E_{\rm in}(\w)
  \end{pmatrix}
  ={\cal U}_{\rm g}  \begin{pmatrix}
    b_{\rm out}(\w) \\
    c_{\rm out}(\w) \\
    L_{\rm out}(\w) \\
    E_{\rm out}(\w)
  \end{pmatrix},
\end{equation}
where the unitary matrix
\begin{align}
  {\cal U}_{\rm g} &=
  \inv{D(\w)} \left( {\begin{pmatrix}
    {\cal U}_{\rm s} (\frac{\Gamma}{2} + i \delta_{\rm a})(\half{\kappa}+i\delta_{\rm c}) & ig\sqrt{\Gamma}\vec{\nu} \\ \\
    ig\sqrt{\Gamma}\vec{\nu}^T & (\frac{\Gamma}{2} - i \delta_{\rm a})(\half{\kappa}+i\delta_{\rm c})
  \end{pmatrix} - g^2 \one} \right), \label{eqn:2lvlio}
\end{align}
with $\delta_{\rm a} = \w-\w_{\rm ge}$, $\delta_{\rm c} = \w-\w_{\rm c}$, ${\cal U}_{\rm s}$ is defined in equation \eqref{aeqn:cavU}, the vector $\vec{\nu}$ reads
\begin{equation}
  \vec{\nu} = \begin{pmatrix}
    \sqrt{\kappa_{\rm b}} \\
    \sqrt{\kappa_{\rm c}} \\
    \sqrt{\kappa_{\rm L}}
  \end{pmatrix}
\end{equation}
and
\beq
D(\w) = (\frac{\Gamma}{2} + i \delta_{\rm a})(\half{\kappa}+i\delta_{\rm c})+g^2.
\eeq
Notice that this set of equations reduces to \eqref{aeqn:emptyio} for the $b$ and $c$ modes, when $g=0$. Finally, we also assume that all output operators ($b_{\rm out},c_{\rm out},L_{\rm out},E_{\rm out}$) commute with each other, which is not true in general. This approximation is required, for an input coherent field in the $b_{\rm in}$ mode to transform into a coherent reflected field in the $b_{\rm out}$ mode and coherent fields in the $c_{\rm out}$, $L_{\rm out}$ and $E_{\rm out}$ modes.

Lastly, as a consistency check of our work, we note that the approximation $\sig_z \approx -\one$ necessarily requires that the atom remains in the ground state throughout its evolution. Solving for $\sig(\w)$, one obtains
\begin{equation}
  \sig(\w) \approx \frac{-ig \sqrt{\kappa_{\rm b}} b_{\rm in}(\w)}{(\frac{\Gamma}{2} - i \delta_{\rm a})(\half{\kappa}-i\delta_{\rm c}) + g^2}, \label{aeqn:sigw}
\end{equation}
where we have used the fact that the $c_{\rm in}$ and $\sig_{\rm in}$ modes represent thermal noise and are thus taken to be negligible compared to the $b_{\rm in}$ mode. Taking the inverse fourier transform, one obtains
\begin{equation}
  P_{\rm e}(t) = \mean{\sig\dagg\sig(t)} = \inv{2\pi} \left| \int \dint\w \, \frac{ig \sqrt{\kappa_{\rm b}} s_{\rm L}(\w)}{(\half{\Gamma}-i(\w-\w_{\rm a}))(\half{\kappa}-i(\w-\w_{\rm c})) + g^2} e^{-i\w t}\right|^2.
\end{equation}
We then conclude that the self-consistency of our approximation requires that $\mean{\sig\dagg\sig(t)} \ll 1$, $\forall t$, which implies that $\max_t P_{\rm e}(t) \ll 1$. Notice that this condition is a necessary but not sufficient condition for the approximation $\sig_z \approx -\one$, since this condition itself is derived under the approximation. However, it should be noted that equation \eqref{aeqn:sigw} can be written in terms of $a(\w)$ to obtain
\begin{align}
  \sig(\w) &\approx \frac{-ig}{\half\Gamma - i (\w-\w_{\rm a})} a(\w). \\
  &= \frac{-ig}{i\Delta(1-\frac{\w-\w_{\rm c}}{\Delta}+\frac{\Gamma}{2i\Delta})} a(\w)
\end{align}
Then, for $\Delta\gg \w-\w_{\rm c}, \Gamma$, meaning that if the laser addresses only wavelengths close to the bare cavity resonance compared to the detuning between the atom and the cavity, and the atom-cavity detuning is many atomic linewidths away from resonance, one has the condition
\begin{equation}
  \mean{\sig \dagg \sig(t)} \approx \pare{\frac{g}{\Delta}}^2 \mean{a\dagg a(t)} \ll 1,
\end{equation}
which is the condition of validity of the dispersive approximation (see \cite{boissonneault2009dispersive}). This means that the condition $\Delta \gg \w-\w_{\rm c},\Gamma$, together with condition $\max_t P_{\rm e}(t) \ll 1$, is sufficient to justify the approximation $\sig_z \approx -\one$.
\subsection*{Derivation of visibility}
The state produced after the cavity and beam splitter is of the form
\begin{equation}
  \cos\nu\ket{s,0}\otimes \ket{\alpha_{\rm o},\alpha_{\rm b},\alpha_{\rm L},\alpha_{\rm E}}+  \sin\nu\ket{g,\tilde{\alpha}}\otimes \ket{\alpha_{\rm o}\,',\alpha_{\rm b}\,',\alpha_{\rm L}\,',\alpha_{\rm E}\,'}
\end{equation}
where o,b,L and E are the other port of the beam splitter, the reflected field from the cavity, the field loss in cavity mirrors and the spontaneous emission term respectively. If we measure only the atomic state and the $a$ mode, we lose information in all the other modes. Tracing over all the rest of the modes, we obtain the state \eqref{eqn:real_state}, where the visibility, $V$ is,
\begin{equation}
  V = |\braket{\alpha_{\rm o}}{\alpha_{\rm o}\,'}\braket{\alpha_{\rm b}}{\alpha_{\rm b}\,'} \braket{\alpha_{\rm L}}{\alpha_{\rm L}\,'}  \braket{\alpha_{\rm E}}{\alpha_{\rm E}\,'}|
\end{equation}
We take only the magnitude of the inner products, since the total phase can, in principle, be compensated by suitable atomic state preparation. This gives the expression
\begin{align}
  V &= \exp\Big[-F \frac{|\tilde{\alpha}|^2}{2 t_{\rm BS}^2} \Big], \label{eqn:V_app} \\
  F&= f_{\rm cav} + I_{s_{\rm L}} \inv{4 C} (1+f_{\rm cav}) + r_{\rm BS}^2, \\
  I_{s_{\rm L}} &= \frac{\int d\w |s_{\rm L}(\w)\inv{D(\w)}|^2}{\int d\w |s_{\rm L}(\w)\inv{D(\w)} \inv{1+i2 (\w-\w_{\rm c})/\kappa}|^2}, \\
  C &= \frac{g^2}{\Gamma \kappa}, \\
  f_{\rm cav} &= \frac{\kappa_{\rm b} + \kappa_{\rm L}}{\kappa_{\rm c}}, \label{eqn:f_cav_app}
\end{align}
where, as in the main text, $t_{\rm BS}$ and $r_{\rm BS}$ are the transmittivity and reflectivity of the beam splitter used in the displacement operation, $s_{\rm L}(\w)$ is the spectrum of the laser input field and $\kappa_i$ is the coupling rate of the cavity mode to the $i$th bath.


\begin{thebibliography}{10}

\bibitem{diqkd}
A.~Ac\'in, N.~Brunner, N.~Gisin, S.~Massar, S.~Pironio, and V.~Scarani,
  ``Device-Independent Security of Quantum Cryptography against Collective
  Attacks,'' \href{http://dx.doi.org/10.1103/PhysRevLett.98.230501}{{\em Phys.
  Rev. Lett.} {\bfseries 98}, 230501 (2007)}.

\bibitem{dirandom}
S.~Pironio, A.~Ac\'in, S.~Massar, A.~B. d.~l. Giroday, D.~N. Matsukevich,
  P.~Maunz, S.~Olmschenk, D.~Hayes, L.~Luo, T.~A. Manning, and C.~Monroe,
  ``Random numbers certified by Bell's theorem,''
  \href{http://dx.doi.org/10.1038/nature09008}{{\em Nature} {\bfseries 464},
  1021--1024 (2010)}.


\bibitem{distateest}
C.-E. Bardyn, T.~C.~H. Liew, S.~Massar, M.~McKague, and V.~Scarani,
  ``Device-independent state estimation based on Bell's inequalities,''
  \href{http://dx.doi.org/10.1103/PhysRevA.80.062327}{{\em Phys. Rev. A}
  {\bfseries 80}, 062327 (2009)}.

\bibitem{rabelo2011device}
R.~Rabelo, M.~Ho, D.~Cavalcanti, N.~Brunner, and V.~Scarani,
  ``Device-Independent Certification of Entangled Measurements,''
  \href{http://dx.doi.org/10.1103/PhysRevLett.107.050502}{{\em Phys. Rev.
  Lett.} {\bfseries 107}, 050502 (2011)}.

 \bibitem{review}N.~Brunner, D.~Cavalcanti, S.~Pironio, V.~Scarani, S.~Wehner. ``Bell nonlocality'', Preprint at http://arxiv.org/abs/1303.2849 (2013).

\bibitem{aspect_experimental_1982}
A.~Aspect, J.~Dalibard, and G.~Roger, ``Experimental Test of Bell's
  Inequalities Using Time- Varying Analyzers,''
  \href{http://dx.doi.org/10.1103/PhysRevLett.49.1804}{{\em Phys. Rev. Lett.}
  {\bfseries 49}, 1804--1807 (1982)}.

\bibitem{weihs_violation_1998}
G.~Weihs, T.~Jennewein, C.~Simon, H.~Weinfurter, and A.~Zeilinger, ``Violation
  of Bell's Inequality under Strict Einstein Locality Conditions,''
  \href{http://dx.doi.org/10.1103/PhysRevLett.81.5039}{{\em Phys. Rev. Lett.}
  {\bfseries 81}, 5039--5043 (1998)}.

\bibitem{scheidl_violation_2010}
T.~Scheidl, R.~Ursin, J.~Kofler, S.~Ramelow, X.~Ma, T.~Herbst, L.~Ratschbacher,
  A.~Fedrizzi, N.~K. Langford, T.~Jennewein, and A.~Zeilinger, ``Violation of
  Local Realism with Freedom of Choice,''
  \href{http://dx.doi.org/10.1073/pnas.1002780107}{{\em PNAS} {\bfseries 107},
  19708--19713 (2010)}.

\bibitem{zeilinger12}
M.~ Giustina, A.~Mech, S.~Ramelow, B.~Wittmann, J.~Kofler, J.~Beyer, A.~Lita, B.~Calkins, T.~Gerrits, S.~W.~Nam, R.~Ursin, A.~Zeilinger, ``Bell violation using entangled photons without the fair-sampling assumption,''
   \href{http://dx.doi.org/10.1038/nature12012}{{\em Nature} (2013)}.

\bibitem{matsukevich_bell_2008}
D.~N. Matsukevich, P.~Maunz, D.~L. Moehring, S.~Olmschenk, and C.~Monroe,
  ``Bell Inequality Violation with Two Remote Atomic Qubits,''
  \href{http://dx.doi.org/10.1103/PhysRevLett.100.150404}{{\em Phys. Rev.
  Lett.} {\bfseries 100}, 150404 (2008)}.

\bibitem{rowe_experimental_2001}
M.~A. Rowe, D.~Kielpinski, V.~Meyer, C.~A. Sackett, W.~M. Itano, C.~Monroe, and
  D.~J. Wineland, ``Experimental violation of a Bell's inequality with
  efficient detection,'' \href{http://dx.doi.org/10.1038/35057215}{{\em Nature}
  {\bfseries 409}, 791--794 (2001)}.

\bibitem{ansmann_violation_2009}
M.~Ansmann, H.~Wang, R.~C. Bialczak, M.~Hofheinz, E.~Lucero, M.~Neeley, A.~D.
  {O'Connell}, D.~Sank, M.~Weides, J.~Wenner, A.~N. Cleland, and J.~M.
  Martinis, ``Violation of Bell's inequality in Josephson phase qubits,''
  \href{http://dx.doi.org/10.1038/nature08363}{{\em Nature} {\bfseries 461},
  504--506 (2009)}.

\bibitem{cavalcanti10}
D.~Cavalcanti, N.~Brunner, P.~Skrzypczyk, A.~Salles, and V.~Scarani, ``Large
  violation of Bell inequalities using both particle and wave measurements,''
  \href{http://dx.doi.org/10.1103/PhysRevA.84.022105}{{\em Phys. Rev. A}
  {\bfseries 84}, 022105 (2011)}.

  \bibitem{quintino12}
M.~{T{\'u}lio Quintino}, M.~{Ara{\'u}jo}, D.~{Cavalcanti}, M.~{Fran{\c c}a Santos}, and
  M.~{Terra Cunha}, ``{Maximal violations and efficiency requirements for Bell
  tests with photodetection and homodyne measurements},''
  \href{http://dx.doi.org/10.1088/1751-8113/45/21/215308}{{\em J.~Phys. A:
  Math. Theor.} {\bfseries 45}, 215308 (2012)}.

\bibitem{araujo11}
M.~{Ara{\'u}jo}, M.~{T{\'u}lio Quintino}, D.~{Cavalcanti}, M.~{Fran{\c c}a
  Santos}, A.~{Cabello}, and M.~{Terra Cunha}, ``{Bell tests with arbitrarily
  low photodetection efficiency and homodyne measurements},''
  \href{http://pra.aps.org/abstract/PRA/v86/i3/e030101}{{\em Phys. Rev.~A}
  {\bfseries 86}, 030101(R) (2012)}.

    \bibitem{sangouard11}
N.~{Sangouard}, J.-D. {Bancal}, N.~{Gisin}, W.~{Rosenfeld}, P.~{Sekatski},
  M.~{Weber}, and H.~{Weinfurter}, ``{Loophole-free Bell test with one atom and
  less than one photon on average},''
  \href{http://dx.doi.org/10.1103/PhysRevA.84.052122}{{\em Phys. Rev.~A}
  {\bfseries 84}, 052122 (2011)}.

\bibitem{chsh69}
J.~F. Clauser, M.~A. Horne, A.~Shimony, and R.~A. Holt, ``Proposed Experiment
  to Test Local Hidden-Variable Theories,''
  \href{http://dx.doi.org/10.1103/PhysRevLett.23.880}{{\em Phys. Rev. Lett.}
  {\bfseries 23}, 880--884 (1969)}.

  \bibitem{brunner07}
N.~Brunner, N.~Gisin, V.~Scarani, and C.~Simon, ``Detection Loophole in
  Asymmetric Bell Experiments,''
  \href{http://dx.doi.org/10.1103/PhysRevLett.98.220403}{{\em Phys. Rev. Lett.}
  {\bfseries 98}, 220403 (2007)}.

\bibitem{cabello07}
A.~Cabello and J.-{\AA}. Larsson, ``Minimum Detection Efficiency for a
  Loophole-Free Atom-Photon Bell Experiment,''
  \href{http://dx.doi.org/10.1103/PhysRevLett.98.220402}{{\em Phys. Rev. Lett.}
  {\bfseries 98}, 220402 (2007)}.

\bibitem{henkel2010highly}
F.~Henkel, M.~Krug, J.~Hofmann, W.~Rosenfeld, M.~Weber, and H.~Weinfurter,
  ``Highly Efficient State-Selective Submicrosecond Photoionization Detection
  of Single Atoms,''
  \href{http://dx.doi.org/10.1103/PhysRevLett.105.253001}{{\em Phys. Rev.
  Lett.} {\bfseries 105}, 253001 (2010)}.

    \bibitem{Brask12}
J. B. Brask, N. Brunner, D. Cavalcanti, and J. Leverrier, ``{The use of amplifiers in continuous-variable Bell tests},''
  \href{http://dx.doi.org/10.1103/PhysRevA.84.052122}{{\em Phys. Rev.~A}
  {\bfseries 85}, 042116 (2012)}.


\bibitem{eberhard_93}
P.~H. Eberhard, ``Background level and counter efficiencies required for a
  loophole-free Einstein-Podolsky-Rosen experiment,''
  \href{http://dx.doi.org/10.1103/PhysRevA.47.R747}{{\em Phys. Rev. A}
  {\bfseries 47}, R747--R750 (1993)}.

\bibitem{larsson01}
J.-\AA. Larsson and J.~Semitecolos, ``Strict detector-efficiency bounds for
  n-site Clauser-Horne inequalities,''
  \href{http://dx.doi.org/10.1103/PhysRevA.63.022117}{{\em Phys. Rev. A}
  {\bfseries 63}, 022117 (2001)}.

\bibitem{matteo_g.a._displacement_1996}
M.~G.~A.~Paris, ``Displacement operator by beam splitter,''
  \href{http://dx.doi.org/10.1016/0375-9601(96)00339-8}{{\em Phys. Lett. A}
  {\bfseries 217}, 78--80 (1996)}.

\bibitem{walls_quantum_2008}
D.~Walls and G.~J. Milburn, {\em Quantum Optics}.
\newblock Springer, 2nd~ed., Feb., 2008.

\bibitem{syed2010}
S.~A.~Aljunid, B.~Chng, M.~Paesold, G.~Maslennikov and C.~Kurtsiefer, ``Interaction of light with a single atom in the strong focusing regime'',
   Preprint at \href{http://arxiv.org/abs/1006.2191}{http://arxiv.org/abs/1006.2191 (2010)}.

\bibitem{morrin1994}
S.~E.~Morrin, C.~C.~Yu and T.~W.~Mossberg, ``Strong Atom-Cavity Coupling over Large Volumes and the Observation of Subnatural Intracacvity Atomic Linewidths'', {\em Phys. Rev. Lett.}, {\bf 73}, 1489 (1994)

\bibitem{birnbaum2005photon}
K.~Birnbaum, A.~Boca, R.~Miller, A.~Boozer, T.~Northup, and H.~Kimble, ``Photon
  blockade in an optical cavity with one trapped atom,''
  \href{http://dx.doi.org/10.1038/nature03804}{{\em Nature} {\bfseries 436},
  87--90 (2005)}.

\bibitem{hood_thesis}
C. J. Hood, ``Real-time Measurement and Trapping of Single Atoms by Single Photons'' PhD dissertation, California Institute of Technology (2000)

\bibitem{ritter_elementary_2012}
S.~Ritter, C.~N\"olleke, C.~Hahn, A.~Reiserer, A.~Neuzner, M.~Uphoff, M.~M\"ucke,
  E.~Figueroa, J.~Bochmann, and G.~Rempe, ``An elementary quantum network of
  single atoms in optical cavities,''
  \href{http://dx.doi.org/10.1038/nature11023}{{\em Nature} {\bfseries 484},
  195--200 (2012)}.



\bibitem{BC12}
J. B. Brask and R. Chaves, ``Robust nonlocality tests with displacement-based measurements'',
  \href{http://pra.aps.org/abstract/PRA/v86/i1/e010103}{{\em Phys. Rev. A}
  {\bfseries 86},  010103(R) (2012)}.

\bibitem{boissonneault2009dispersive}
M.~Boissonneault, J.~M. Gambetta, and A.~Blais, ``Dispersive regime of circuit
  QED: Photon-dependent qubit dephasing and relaxation rates,''
  \href{http://dx.doi.org/10.1103/PhysRevA.79.013819}{{\em Phys. Rev. A}
  {\bfseries 79}, 013819 (2009)}.




\end{thebibliography}

\section*{Acknowledgements}

We would like to thank Ignacio Cirac for pointing out the issue of cooperativity and also Gerhard Rempe, Stephan Ritter, Serge Haroche, Christian Kurtsiefer, Gleb Maslennikov, Matthias U. Staudt, Antonio Badolato, Dario Gerace and Steve Girvin for useful discussions. This work was supported by the Brazilian agencies Fapemig, Capes, CNPq, and INCT-IQ, the National Research Foundation and the Ministry of Education of Singapore.

\section*{Authors contributions} \dani{M. Ara\'ujo, M. T. Quintino, D. Cavalcanti, M. T. Cunha, and M. F. Santos worked on the ideal case and Bell test proposal.  C. Teo, J. Min\'{a}\v{r}, M. F. Santos, and V. Scarani, developed the calculations for the realistic case. All authors discussed the results and wrote the manuscript.}

\section*{Competing financial interests} The authors declare no competing financial interests.

\end{document}